\documentclass[5p]{elsarticle}
\usepackage{hyperref} 
\usepackage{textcomp}
\usepackage{amssymb}
\usepackage{amsmath}

\begin{document}

\begin{frontmatter}

\title{Active correction of the tilt angle of the surface plane with respect to the rotation axis during azimuthal scan}

\author{M. Sereno, S. Lupone, M. Debiossac, N. Kalashnyk, P.Roncin}
\address{Institut des sciences mol\'{e}culaires d'Orsay (ISMO), CNRS, Univ. Paris-Sud, Universit\'{e} Paris-Saclay, Orsay F-91405, France}

\begin{abstract}
A procedure to measure the residual tilt angle $\tau$ between a flat surface and the azimuthal rotation axis of the sample holder is described. When the incidence angle $\theta$ and readout of the azimuthal angle $\phi$ are controlled by motors, an active compensation mechanism can be implemented to reduce the effect of the tilt angle during azimuthal motion. After this correction, the effective angle of incidence is kept fixed, and only a small residual oscillation of the scattering plane remains. \end{abstract}


\end{frontmatter}


\section{Introduction} 
Surface studies using grazing incidence angles offer a high degree of surface sensitivity. This the case for X-rays\cite{Renaud}, high energy electrons\cite{Pukite} and ions\cite{Pfandzelter} as well as fast atoms \cite{Roncin2002,Seifert_in_situ}. In these techniques the angle of incidence is only few degrees. For most diffraction experiments, the required high positioning accuracy of the sample is provided with a goniometer but this is hardly compatible with sample transfer under UHV conditions and with high temperature needed for annealing. The real challenge appears when the structural information lies in an intensity variation during an azimuthal rotation of the surface such as in ion beam triangulation \cite{Pfandzelter} and even more drastically in atom beam triangulation \cite{Seifert_alanine,Nataliya}. Ensuring stability of the incidence angle better than 0.1 ° requires active correction. We describe here a method designed for atom beam triangulation in a grazing incidence fast atom diffraction setup (GIFAD) that solves this issue.
\begin{figure}[ht]
	\begin{center}\includegraphics [width=70mm]{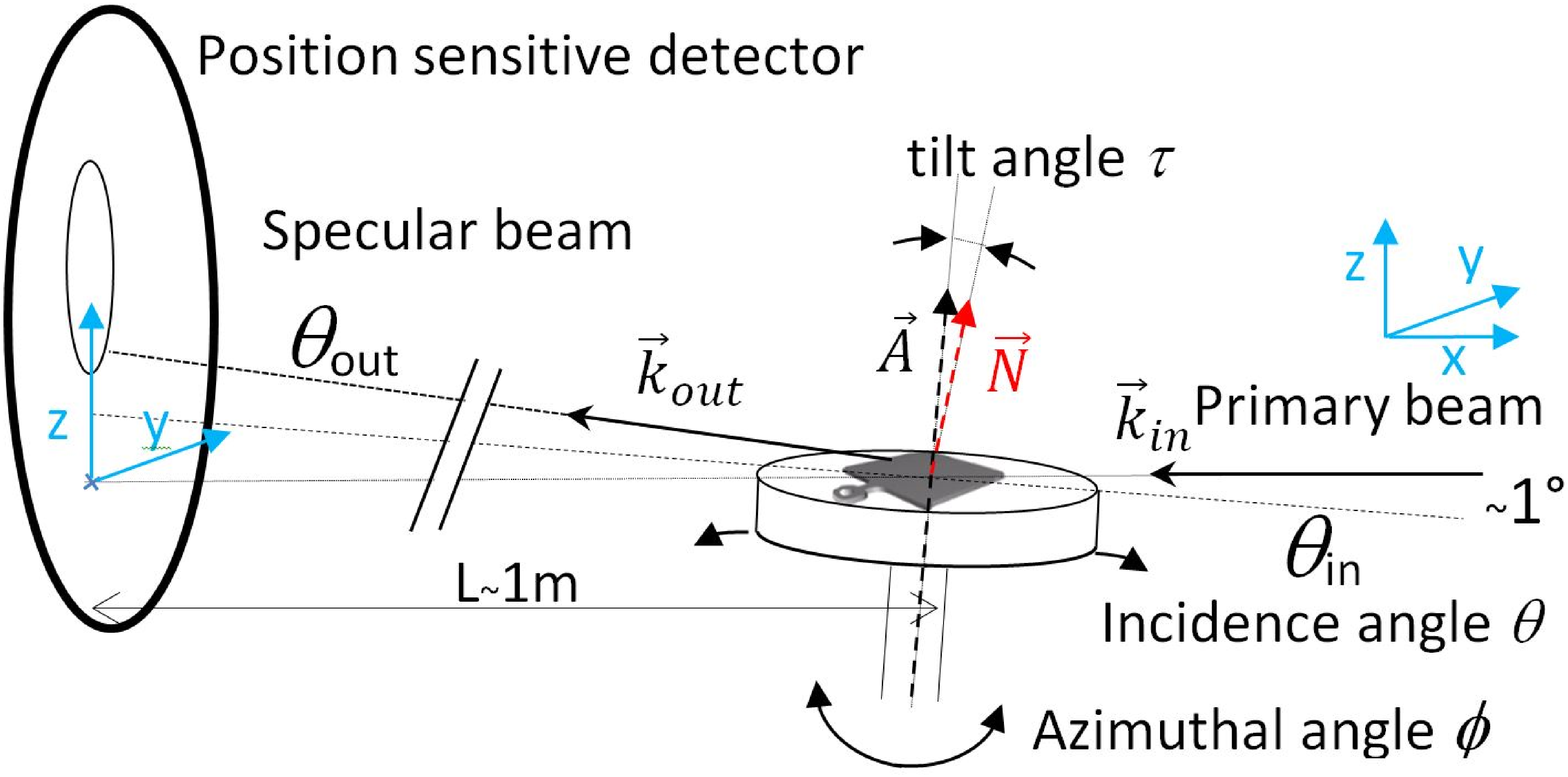}\end{center}
	\caption{{Typical grazing incidence setup where the incident beam has a fixed direction. If the surface plane, defined here by its normal $\vec{N}$, is not exactly perpendicular to the rotation axis $\vec{A}$, the specular beam follows characteristic orbits during azimuthal scan.}}
	\label{schematic}
\end{figure}
\section{Grazing incidence fast atom diffraction}
Grazing incidence fast atom diffraction (GIFAD or FAD) has recently emerged as a new surface science technique with exclusive surface sensitivity and high structural resolution \cite{Seifert_SiO2,Debiossac2}. In this technique a beam of $\sim$keV atoms impinges the surface at incidence angles $\theta$ of about one degree and with a angular spread (divergence) typically around 0.01 degree. When the beam is aligned with a low index crystal direction, diffraction can take place provided that the surface quality is good enough in terms of flatness and coherence length. The latter quantity is defined here as the mean separation between crystal defects. First observed on single crystals of wide band gap ionic insulators such as LiF \cite{Schuller_2007,Rousseau_2007}, NaCl \cite{Lalmi}, GIFAD has been shown to be successful in measuring metal surfaces \cite{Bundaleski,Schuller_2009}, semi-conductors \cite{Khemliche,Debiossac2} and even single layers of inorganic compounds \cite{Seifert_SiO2} and organic molecules \cite{Seifert_alanine,Busch}. For surfaces of organic molecules, diffraction is not as sharp as for bulk crystal surfaces or inorganic layers probably due to the presence of a higher density of structural defects. Fortunately, the triangulation technique \cite{Pfandzelter,Seifert_alanine,Nataliya} easily reveals the alignment direction of the adsorbed molecules while the shape of the azimuthal profile gives hints on the detailed molecular assembly \cite{Seifert_alanine}. Even faint diffraction is enough to derive the lattice parameter of the molecular organization \cite{Seifert_alanine}. Also, it has been suggested \cite{Debiossac_PRA,Zugarramurdi} that azimuthal scans around channeling directions could be use to measure the range $R_c$ of the interaction potential above the surface. This fundamental parameter can be very useful for quantitative analysis since it governs the length $L_T\sim R_c /\theta$ of the atom trajectory above the surface\cite{Debiossac_PRA}. At high energies, selective sputtering of step edges is also sensitive to the crystallographic direction \cite{Michely}. In addition, GIFAD has demonstrated  pronounced intensity oscillations during layer-by-layer molecular beam epitaxy growth\cite{Atkinson}. Following these oscillations during azimuthal rotation would allow a better uniformity of the layers.
Ideally, these azimuthal scans should be performed without changing the angle of incidence. This turns out to be more difficult than expected. On one hand it is not straightforward to put a goniometer under UHV condition. On the other hand, sample transfer and rotation devices do not always allow for ultra-precise positioning of the sample. Most often the surface normal has a small residual tilt angle $\tau$ of about one degree with respect to the rotation axis. This is sufficient to prevent straight-forward application of purely azimuthal scan. When the control of both the azimuthal and incidence angle is motorized, an on-line correction can be performed and is described below.



\section{The orbit of the specular spot}
For each azimuthal angle, the plane of incidence is defined as the plane containing both the incident beam and the surface normal. It also contains the specularly reflected beam, so that the location of the specular beam on the detector indicates the direction of the surface normal. During an azimuthal scan the specular beam spot follows a curve that can be described as an orbit. Using the standard description of specular scattering.
\begin{equation} \label{eq1}  
\overrightarrow{k_{spec}} = \overrightarrow{k_{in}} - 2(\overrightarrow{k_{in}}.\overrightarrow{N}) \overrightarrow{N} \end{equation}

Where $ \overrightarrow{k_{in}}=-k_0 \vec{x}$ is the incoming wavevector, $\overrightarrow{k_{spec}}$ is the specular one and $\overrightarrow{N}$ is the surface normal. The exact calculation is not difficult but becomes even simpler by taking into account that the detector is located far away from the target surface so that the variation of the location of the impact on the crystal surface  can be neglected. 
 We consider here that the beam impacts the surface exactly on its intersection with the rotation axis. Accordingly, the coordinates (y,z) on the detector depend only on the scattering angles. 
 We only need to focus on the vector $\vec{N}(\phi)$ describing the surface normal tilted by an angle $\tau$ with respect to the rotation axis $\vec{A}$ during azimuthal rotation $\phi$ around this axis. Defining the rotation around a vector $\vec{u}$ by an angle $\alpha$ as  $R_{\vec{u}}(\alpha)$, $\overrightarrow{N}(\phi)$  can be written as a product of three rotations describing the tilt angle angle $R_{\vec{y}}(\tau)$, the azimuthal scan $R_{\vec{z}}(\phi)$ and the angle of incidence $\theta_0$, $ R_{\vec{y}}(\theta_0)$; 
 
\[ \overrightarrow{N}(\phi) =R_{\vec{y}}(\theta_0)R_{\vec{z}}(\phi)R_{\vec{y}}(\tau) \vec{z} \]
\[  \overrightarrow{N}(\phi) = \begin{bmatrix}
\cos\tau\sin\theta_0 + \cos\phi\cos\theta_0\sin\tau \\
\sin\phi\sin\tau \\
\cos\tau\cos\theta_0 - \cos\phi\sin\tau\sin\theta_0\\
\end{bmatrix}
\]

Considering for simplicity, unit vectors $\overrightarrow{k_{in}}$ and $\overrightarrow{k_{out}}$ , i.e. $k_0=1$,  the scalar product $\overrightarrow{k_{in}}.\overrightarrow{N}$ is the first coordinate of the vector $\vec{N}$ so that, within eq.\ref{eq1}, the $y$ and $z$ components of the specular wave-vector write;
		\[k_y(\phi) = -2\sin\phi\sin\tau(\cos\theta_0\cos\phi\sin\tau-\sin\theta_0\cos\tau)\] 
		\[k_z(\phi) = -2\cos\tau(\cos\phi\cos\theta_0\sin\tau-\sin\theta_0\cos\tau)\] 

with $\phi$ the azimuthal angle. For small angle $\theta_0$ and $\tau$ the formula simplify and the coordinates $y(\phi), z(\phi)$ on the detector form the parametric equations of the orbit.
\begin{equation} \label{eq2}
	\begin{split}   
		y(\phi) = -2L\tau\sin\phi(\tau\cos\phi-\theta_0)\\
		z(\phi) = 2L\theta_0-2L\tau\cos\phi\\
	\end{split}
\end{equation}

\begin{figure}[ht]
		\begin{center}\includegraphics [width=80mm]{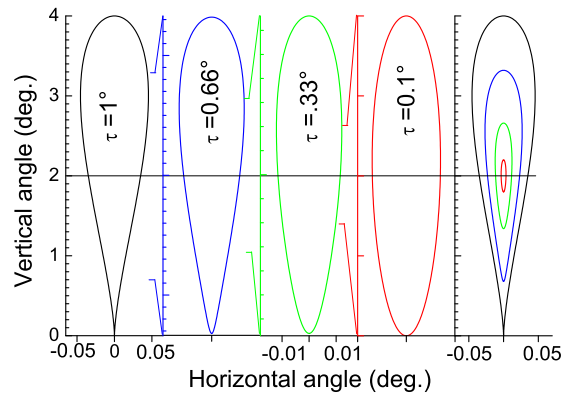}\end{center}
	\caption{{Orbits corresponding to an angle of incidence $\theta_0$ =1$^\circ$ and to a tilt angle $\tau$ of 1$^\circ$, 0.66$^\circ$, 0.33$^\circ$ and 0.1$^\circ$. For each panel, the horizontal scale is zoomed $\sim$ ten times compared with the vertical one. The rightmost panel plots the four orbits at the same scale.}}
	\label{iresp}
\end{figure}
The eq.\ref{eq2} shows that the full vertical extension $W_z=4L\tau$ and full width at half height $W_y=4L\tau\theta_0$ are simple quantities while figure \ref{iresp} illustrates that the orbits are not simple ellipses.
\begin{figure}[ht]
	\begin{center}\includegraphics [width=60mm]{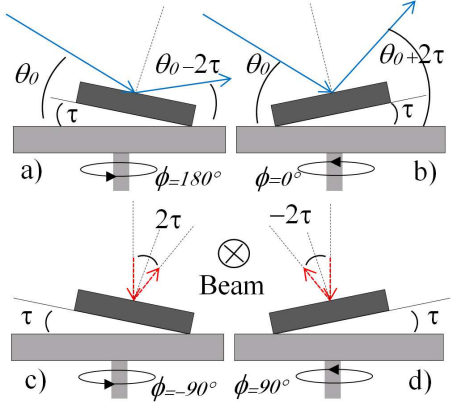}\end{center}
	\caption{{The plane containing the rotation axis and the surface normal is plotted for four different azimuth angles $\phi$. Twice per revolution, at angles (b) $\phi=0^\circ$ and (a) $\phi=180^\circ$, this plane contains the primary beam, and coincides with the incidence plane. The outgoing angles differ by $4\tau$ and the associated spots are separated by $W_z$. At 90$^\circ$ from these values, (c) and (d), the beam momentum $\vec{k}$ is almost perpendicular to the reference plane with only a tiny component $\vec{k}\sin\theta_0$ parallel to the rotation axis (in dashed red). In this projected plane, the variation in the deflection of the specular beam is again 4$\tau$ but the amplitude on the detector will only be $W_y=W_z\sin\theta_0$, i.e. almost two order of magnitude smaller than $W_z$.}}
	\label{tilt}
\end{figure}

\section{Experimental measure of $\tau$}
\begin{figure}[ht]
	\begin{center}\includegraphics [width=60mm]{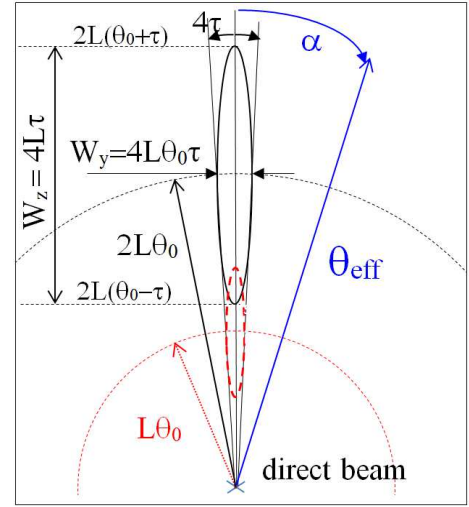}\end{center}
	\caption{{Schematic view of the orbit on the detector with the polar coordinates ($\theta_{eff},\alpha$). Taking the direct beam location as a reference $\theta_{eff}=\rho/L$ and $\alpha=atan(y/\rho)$ with  $\rho=\sqrt{y^2+z^2}$ and \textit{L} the distance from the surface to the detector. The small angle approximation is used, i.e. $sin\theta_0\sim\theta_0$ and $sin\tau\sim\tau$. The orbit in dashed red corresponds to the orbit of the line defined as the intercept of the incidence plane with the surface.}}
	\label{scheme_2}
\end{figure}

The tilt angle has to be measured in situ because each transfer can introduce slight variation of this value. The simplest reference that can be measured with high accuracy is the location of the primary beam on the detector. This is achieved by removing the target from the beam for a fraction of a second. From eq. \ref{eq2} the  tilt angle $\tau$ can be calculated as the ratio of $W_z$ to $W_y$. Another option is to measure accurately several values to fit the parametric curve onto these data points. Intuitively, in the elliptic case, five values are needed corresponding to the coordinates $y,z$ of the two focal points of the ellipse and to the length of the chord. These are  linked to the experimental parameters : the value of the tilt angle $\tau$, the mean scattering angle $\theta_0$, the direct beam position and the mean inclination of the scattering plane $\alpha_0$. 
As depicted in fig.\ref{scheme_2} these values can be determined as simple ratios between geometric distances measured on the detector. In particular the mean scattering angle $\theta_0 $ can be estimated as $\theta_0=W_y/W_z$ or via the center of the ellipse $\bar{z}=(z_{max}+z_{min})/2=2L\theta_0$ while $\tau$ can be estimated as $\tau=W_z/4L$ or $\tau=W_y/4L\theta_0$. 
The main difficulty is that, quite often, a new surface has so many defects that the scattering profile of the helium atoms can be very diffuse with full width half maximum (FWHM) $\sigma_y$ and $\sigma_z$ such that the associated angular widths $\sigma_\alpha = \sigma_y/L$ and $\sigma_\theta = \sigma_z /L$ exceed one degree. A possible solution is to track the mean value of the scattering distribution either by moment analysis or via fit procedures which can usually bring the statistical error down to $\sigma/10$ or even $\sigma/100$ if the number of data points is large enough \cite{Nataliya}. When the tilt is larger than the angle of incidence, closed orbit does not exist because the reflection is not possible in a selected azimuthal range. Even in this case, the reduced variation of the scattering plane can be measured by tracking the mean value of the scattering profile along \textit{y}.
\begin{figure}[ht]
	\begin{center}\includegraphics [width=60mm]{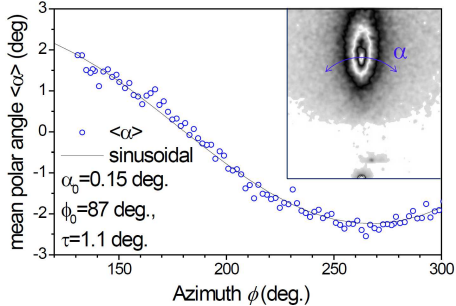}\end{center}	
	\caption{{During an azimuthal scan, the mean polar angle $\langle\alpha\rangle$ of the scattering distribution (ex. in insert) is reported as a function of the azimuth $\phi$ follows a smooth sinusoidal curves having a full amplitude of 4$\tau$. Note that the quasi elliptic shape of the scattering profile is accidental and has nothing to do with the orbits discussed above.}}
	\label{swing}
\end{figure}
Such a condition is illustrated in the insert of Fig. \ref{swing} recorded at an incidence angle $\theta_0$ = 0.7$^\circ$. Surprisingly, a tilt angle of $1.1^\circ$ that is almost two times larger than the angle of incidence could be measured. In this case the azimuthal scan was impossible without adjusting manually the incidence angle when it approaches the value near 0° (almost parallel to the surface). Therefore, the angle of incidence is discontinuous and no orbit can be drawn. As a consequence, the scattering pattern is not analyzed in the Cartesian coordinates ($k_y,k_z$) but in the polar coordinates ($\theta_{eff},\alpha$) illustrated in Fig. \ref{scheme_2}. The smooth oscillation of the collision plane follows a sinusoidal curve indicating  an inclination $\alpha_0$ = 0.15$^\circ$ of the position sensitive detector with respect to the mean scattering plane, a tilt angle $\tau= 1.1^\circ$. 

\section{Test bench with laser pointer}

To test the software of the correction procedure while avoiding possible damage inside the UHV chamber during test, a simple system outside vacuum has been built. A small manipulator holds a laser pointer with its beam  directed onto a polished silicon wafer placed on top of a rotating stepper motor. A second miniature linear stepper motor is used to control the incidence angle $\theta$. The decisive advantage is that at the micron scale most surfaces are quasi perfect so that the specularly reflected laser beam is a single spot that can be imaged onto a paper screen at a distance \textit{L} a few meters downstream. Another advantage is that the impact on the surface is visible making adjustments straightforward. 

The motors are controlled by an inexpensive micro-controller \textgravedbl Arduino Mega \textacutedbl board with a 3D printer \textgravedbl RepRap 1.4\textacutedbl board capable of hosting up to five "Allegro A4988" stepper motor controllers. The Arduino is connected to the host PC via an USB-Link through which it receives order and transmit results. The program is written in C inside the open source, integrated Arduino development environment hosted in a PC and uploaded to the micro-controller by the USB-Link. Various ratios of tilt angle to incidence angle have been explored. In these test conditions the full trajectory can be recorded in few seconds. We found that the curvature radius is so large compared with the horizontal width that the easiest starting strategy is to consider the orbit as a quasi ellipse. The center $(\bar{y},\bar{z})$ of the orbit as well as its inclination $\bar{\alpha}$ are measured directly while the mean scattering angle $\theta_0$ and the tilt angle $\tau$ are determined from the amplitudes $W_y$ (at half height) and $W_z$.

\section{In situ laser reflection}
When the sample is placed in-between two view-ports located opposite to each other on
the UHV chamber, shining a laser at the surface and observing the reflected beam is as easy as on the test bench. A simple screen a few meters away will convert mm resolution to $1/100$ of a degree. A single azimuthal rotation, even partial, is sufficient to measure the tilt angle and the azimuthal reference. If the relative positions of the laser and atom beam axis are known, the azimuthal reference can be transferred to the atom beam.

\section{Correction of incidence angle}
\begin{figure}[ht]
	\begin{center}\includegraphics [width=70mm]{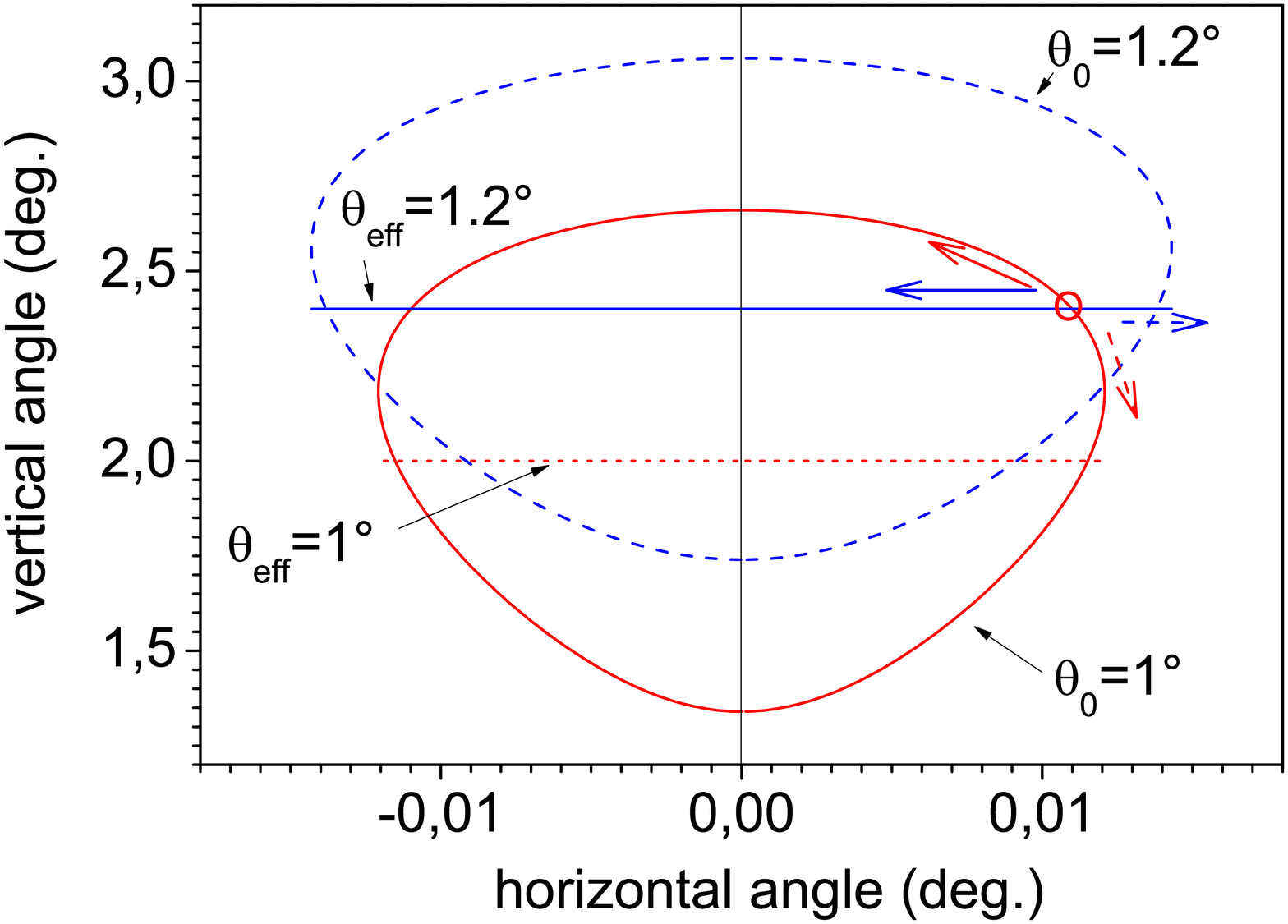}\end{center}
	\caption{{Illustration of the correction. The system is evolving along the red orbit corresponding here to a tilt angle of 0.33$^\circ$ and a mean angle of incidence $\theta_0=1^\circ$. When correction is required by the user, the system is at the location indicated by a circle corresponding to an effective scattering angle of 1.2$^\circ$. The system is now expected to evolve along the corrected circle (looking here as a straight segment at this scale) corresponding to the dashed blue orbit associated with a mean incidence angle $\theta_0=1.2^\circ$.}}
	\label{correction}
\end{figure}

From Eq. \ref{eq2}, the correction to apply to the angle of incidence during rotation is a simple sinusoidal function having a full amplitude of 4$\tau$ and a phase reference $\phi_0$. These parameters are independent of the incidence angle and have to be measured for each new sample surface before applying any correction. The only practical issue is the user interface. To avoid possible problems during sample transfer or large amplitude movement, the default settings are that motor actions are performed without corrections.
When an azimuthal scan is to be programmed, the azimuthal angle $\phi_{start}$ is, a priori, at a random position with respect to the reference plane. At this position, the mean scattering angle $\theta_0$ defined above, i.e. as the average incidence during an uncorrected azimuthal scan, has no particular meaning. At variance, the user should simply assume that the correction will maintain the effective angle of incidence $\theta_{eff}=\theta_{start}=\theta_0+\tau\cos\phi_{start}$ at the moment where the azimuthal scan is decided. It means that the system should evolve on a quasi perfect straight line associated with the incidence angle $\theta_{eff}$ and not along the correction contour associated with the mean value $\theta_0$ of the initial orbit. Care has to be taken to adjust the phase with the proper sign as illustrated in Fig.\ref{correction}. Note that the corrected orbit is now almost a perfect line with less than $10^{-3}$ degree vertical amplitude. The compensation software have been tested successfully on the test bench but, due to a problem with the manipulator, we have no results for atoms scattering inside UHV.

\section{Conclusion}

It should be noticed that the specular angle can be ill-defined. This occurs, for instance, in the case when surface twist and tilt mosaicity are present \cite{Lalmi}. In this case there are several specular beams corresponding to the local surface orientation illuminated by the primary beam so that the correction can not be well defined. With a single crystal surface correction is straightforward but is severely limited by the quality of the mechanical transmission. This includes the difficult problem of mechanical backlash of the vacuum transmission because the correction applied to the angle of incidence has to be reversed at least once per turn. There is still the possibility of a 180$^\circ$ azimuthal scan without backlash. In practice, the presence of two view-ports allowing optical measurement is certainly a favorable condition to measure the tilt angle rapidly and accurately.

\section{Acknowledgment}
We are grateful to Paola Atkinson for help in reading the manuscript, to Ala Husseen for help during the measurements and to Christophe Charri\`{e}re for his assistance in implementing the correction in the user interface software.


\section{References}
\begin{thebibliography}{5}
    
\bibitem{Renaud} G. Renaud, R. Lazzari, F. Leroy, Probing surface and interface morphology with Grazing Incidence Small Angle X-Ray Scattering. Surface Science Reports \textbf{64}, p 255–380 (2009) doi:10.1016/j.surfrep.2009.07.002
    
\bibitem{Pukite} J.M. Van Hove, P.R. Pukite, P.I. Cohen, Reflection high energy electron diffraction measurements of AlGaAs growth instabilities and roughening rates on misoriented substrates. Journal of Vacuum Science and Technology B {\bf3} (2), 563-567 (1987).
    
\bibitem{Pfandzelter} R. Pfandzelter, T. Bernhard, and H. Winter, Ion Beam Triangulation of Ultrathin Mn and CoMn Films Grown on Cu(001). Phys. Rev. Lett. {\bf90}, 036102 ( 2003).

\bibitem{Roncin2002} P. Roncin, A. G. Borisov, H. Khemliche, A. Momeni, A. Mertens, and H. Winter. Evidence for $F^-$ Formation by Simultaneous Double-Electron Capture during Scattering of $F^+$ from a LiF(001) Surface. Phys. Rev. Lett. \textbf{89}, 043201 (2002)
     
\bibitem{Seifert_alanine} J. Seifert, M. Busch, E. Meyer, and H. Winter, Surface structure of alanine on Cu(110) via grazing scattering of fast atoms and molecules.Phys. Rev. B \textbf{89}, 075404 (2014)

\bibitem{Nataliya} N. Kalashnyk, H. Khemliche and P. Roncin. Atom beam triangulation of organic layers at 100 meV normal energy: self-assembled perylene on Ag(1 1 0) at room temperature. Applied Surface Science. \textbf{364}, p 235–240 doi:10.1016/j.apsusc.2015.12.134 (2015).
   
\bibitem {Seifert_in_situ} J. Seifert, H. Winter, In-situ monitoring of oxygen adsorption at Mo(112) surface via fast atom diffraction. Surface Science Vol. 610 L1 (2013).
              
\bibitem {Seifert_SiO2} J. Seifert, A. Sch\"{u}ller, H. Winter, R. W\l{}odarczyk, J. Sauer, and M. Sierka, Diffraction of fast atoms during grazing scattering from the surface of an ultrathin silica film on Mo(112). Phys. Rev. B {\bf82}, 035436 (2010).
     
\bibitem{Debiossac2} M. Debiossac, A. Zugarramurdi, H. Khemliche, P. Roncin, A. G. Borisov, A. Momeni, P. Atkinson, M. Eddrief, F. Finocchi, and V. H. Etgens, Combined experimental and theoretical study of fast atom diffraction on the $\beta_2$(2×4) reconstructed GaAs(001) surface. Phys. Rev. B {\bf90}, 155308 (2014). 

\bibitem{Schuller_2007} A. Sch\"{u}ller, S. Wethekam, and H. Winter, Diffraction of Fast Atomic Projectiles during Grazing Scattering from a LiF(001) Surface. Phys. Rev. Lett. {\bf98}, 016103 (2007).

\bibitem{Rousseau_2007} P. Rousseau, H. Khemliche, A. G. Borisov, and P. Roncin, Quantum Scattering of Fast Atoms and Molecules on Surfaces. Phys. Rev. Lett. {\bf98}, 016104 (2007).

\bibitem{Lalmi} B. Lalmi, H. Khemliche, A. Momeni, P. Soulisse and P. Roncin, High resolution imaging of superficial mosaicity in single crystals using grazing incidence fast atom diffraction. J. Phys.: Condens. Matter {\bf24} 442002 (2012).

\bibitem{Bundaleski} N. Bundaleski, H. Khemliche, P. Soulisse, and P. Roncin, Grazing Incidence Diffraction of keV Helium Atoms on a Ag(110) Surface. Phys. Rev. Lett. {\bf101}, 177601 (2008).

\bibitem{Schuller_2009} A. Sch\"{u}ller, M. Busch, S. Wethekam, and H. Winter, Fast Atom Diffraction from Superstructures on a Fe(110) Surface. Phys. Rev. Lett. {\bf102}, 017602 (2009). 

\bibitem{Khemliche} H. Khemliche, P. Rousseau, P. Roncin, V. H. Etgens and F. Finocchi, Grazing incidence fast atom diffraction: An innovative approach to surface structure analysis. Appl. Phys. Lett. {\bf95}, 151901 (2009).
    
\bibitem{Busch} J. Seifert, M. Busch, E. Meyer, and H. Winter, Surface Structure of Alanine on Cu(110) Studied by Fast Atom Diffraction. Phys. Rev. Lett. {\bf111}, 137601 (2013)..

	
\bibitem{Debiossac_PRA}	M. Debiossac and P. Roncin, Atomic diffraction under oblique incidence: An analytical expression, Phys. Rev. A {\bf90}, 054701 (2014).
   
\bibitem{Zugarramurdi} A. Zugarramurdi and A.G. Borisov, Theoretical study of the effect of beam misalignment in fast-atom diffraction at surfaces.  Phys. Rev. A {\bf87}, 062902 (2013).

\bibitem{Michely} H. Hansen, C. Polop, T. Michely, A. Friedrich, and H.M. Urbassek, Step Edge Sputtering Yield at Grazing Incidence Ion Bombardment.  Phys. Rev. Lett. {\bf92}, 246106 (2004).
	
\bibitem {Atkinson} P. Atkinson, M. Eddrief, V. H. Etgens, H. Khemliche, M. Debiossac, A. Momeni, M. Mulier, B. Lalmi, and P. Roncin,Dynamic grazing incidence fast atom diffraction during molecular beam epitaxial growth of GaAs. Applied Physics Letters {\bf105}, 021602 (2014).

	
\end{thebibliography}
\end{document}